# Signatures of Classical Periodic Orbits on a Smooth Quantum System

Daniel Provost*

*Center for Theoretical Physics, Laboratory for Nuclear Science, and Department of Physics, Massachusetts Institute of Technology, Cambridge, Massachusetts 02139 U.S.A.*

*and*

*Department of Physics and Astronomy, Laurentian University, Sudbury, Canada P3E 2C6*

Gutzwiller's trace formula and Bogomolny's formula are applied to a non–specific, non–scalable Hamiltonian system, a two–dimensional anharmonic oscillator. These semiclassical theories reproduce well the exact quantal results over a large spatial and energy range.

PACS numbers: 05.45.+b, 03.65.Sq, 03.65.Ge

## I. INTRODUCTION

For bound conservative Hamiltonian systems, the main goal of semiclassical mechanics is to obtain estimates for the quantum eigenvalues and eigenfunctions using classical trajectories. Since these eigenvalues and eigenfunctions are time independent quantities, it is natural to expect that only classical trajectories that define time invariant classical manifolds be important semiclassically. For example, the number of Planck cells in the phase space volume enclosed by the energy shell of energy $E$ gives a good semiclassical estimate for the number of eigenstates with energy less than $E$.

For chaotic systems, the only time invariant manifolds other than the energy shell are periodic orbits. Consequently periodic orbits are at the center of efforts to obtain viable semiclassical quantization schemes. Gutzwiller's trace formula [1] uses periodic orbits to explain the oscillations in the density of states about the average density of states. A similar expression was recently developed by Bogomolny [2] to explain the accentuations and attenuations in the average coordinate probability density about the classical periodic orbits, the so–called *scars* [3].

In this paper we present applications of Gutzwiller's trace formula and Bogomolny's formula to the following generic Hamiltonian system [4]:

$$H = \frac{p_x^2}{2} + \frac{p_y^2}{2} + \frac{\mu}{2}x^2 + \left(y - \frac{x^2}{2}\right)^2 \quad (1)$$

with $\mu = 0.1$. This anharmonic Hamiltonian system has been code named Nelson [5] and has been studied extensively both classically and semiclassically [4–6]. The slow progress in the application of periodic orbit theories to physical systems such as the Nelson Hamiltonian is due to the daunting task of finding all the periodic orbits with periods $\tau$ less than $\tau_{max} = 2\pi\hbar/\epsilon$, where $\epsilon$ is the energy resolution desired. We used a method which guarantees that all periodic orbits up to a chosen maximum period are found. This allowed us to make stringent tests of Gutzwiller's trace formula and of Bogomolny's formula.

The outline of the paper is as follows. In section II we review the semiclassical theory, paying particular attention to symmetry considerations. We present in section III the method we used to find classical periodic orbits of a chaotic system in two spatial dimensions and then go on to find the desired orbits for the Nelson Hamiltonian. In section III B we use these periodic orbits to calculate the semiclassical oscillations in the density of states using Gutzwiller's trace formula for different energy smoothing parameters $\epsilon$. We then compare our results with the exact quantum oscillations in the density of states. Finally we present comparisons of the semiclassical theory for the oscillating part of the energy smoothed probability density with the exact quantum results.

## II. THEORY

Consider a bound Hamiltonian system in two spatial dimensions of the form

$$H(\hat{\mathbf{q}}, \hat{\mathbf{p}}) = \frac{\hat{\mathbf{p}}^2}{2} + V(\hat{\mathbf{q}}). \quad (2)$$

Its spectral probability density $\Delta(\mathbf{q}, E)$ is defined as [2,7]:

$$\Delta(\mathbf{q}, E) = \sum_k \delta(E - E_k) |\langle \mathbf{q}|\psi_k\rangle|^2 \quad (3)$$

where $E_k$ are the eigenvalues of $\hat{H}$ and $|\psi_k\rangle$ are the corresponding eigenstates. The density of states $d(E)$ is obtained by integrating the spectral probability density over the spatial coordinates $\mathbf{q}$:

$$\begin{aligned}d(E) &= \int d\mathbf{q}\ \Delta(\mathbf{q}, E) \\ &= \sum_k \delta(E - E_k). \end{aligned} \quad (4)$$

---

*Present address: Department of Physics and Astronomy, Laurentian University, Sudbury, Canada P3E 2C6.
E–mail: dprovost@neutrino.phys.laurentian.ca

A semiclassical formula for $\Delta(\mathbf{q}, E)$ is obtained by rewriting Eq. (3) as:

$$\Delta(\mathbf{q}, E) = \frac{1}{\pi\hbar} \text{Re} \int_0^\infty dt \langle \mathbf{q}|e^{-i\hat{H}t/\hbar}|\mathbf{q}\rangle e^{iEt/\hbar}. \quad (5)$$

An appropriate semiclassical approximation for the propagator $\langle \mathbf{q}|e^{-i\hat{H}t/\hbar}|\mathbf{q}\rangle$ is then used and the resulting time integral is evaluated by stationary phase. The dominant contribution to this integral comes from "zero length" trajectories, those whose propagation time is infinitesimal; this gives rise to the average spectral probability density $\Delta_{avg}(\mathbf{q}, E)$. The remaining contributions to the time integral come from classical trajectories which start at $\mathbf{q}$ and come back to $\mathbf{q}$ in a nonzero time $t$; these trajectories produce the oscillatory part $\Delta_{osc}(\mathbf{q}, E)$ of the spectral probability density (the so called "scars").

### A. Average spectral probability density

The average spectral probability density $\Delta_{avg}(\mathbf{q}, E)$ is given by

$$\Delta_{avg}(\mathbf{q}, E) = \frac{1}{\pi\hbar} \text{Re} \int_0^\tau dt \ \langle \mathbf{q}|e^{-i\hat{H}t/\hbar}|\mathbf{q}\rangle e^{iEt/\hbar} \quad (6)$$

where $\tau$ denotes a time much smaller than the period of the shortest periodic orbit. The Wigner representation [8–10] for the evolution operator is used to evaluate $\Delta_{avg}(\mathbf{q}, E)$:

$$\Delta_{avg}(\mathbf{q}, E) = \frac{2}{(2\pi\hbar)^3} \text{Re} \int_0^\tau dt \int d\mathbf{p} \ [e^{-i\hat{H}t/\hbar}]_W \, e^{iEt/\hbar} \quad (7)$$

where $[\ldots]_W$ denotes the Weyl symbol of the enclosed operator. For short times [8,10]

$$[e^{-\beta\hat{H}}]_W = e^{-\beta[\hat{H}]_W} \left\{ 1 - \frac{\hbar^2}{8} A_1(\mathbf{q}, \mathbf{p}, \beta) + O(\hbar^4) \right\} \quad (8)$$

where $\beta \equiv it/\hbar$. We shall neglect terms of $O(\hbar^4)$. The Weyl symbol for Hamiltonians of the form (2) is simply $H(\mathbf{q}, \mathbf{p})$. Also [8]

$$A_1(\mathbf{q}, \mathbf{p}, \beta) = \beta^2 \nabla^2 V - \frac{\beta^3}{3}[(\nabla V)^2 + (\mathbf{p} \cdot \nabla)^2 V]. \quad (9)$$

The $p_x p_y$ term in Eq. (9) does not contribute to $\Delta_{avg}(\mathbf{q}, E)$. An integration by parts of the terms containing $p_x^2$ or $p_y^2$ gives

$$\Delta_{avg}(\mathbf{q}, E) \simeq \frac{2}{(2\pi\hbar)^3} \text{Re} \int_0^\tau dt \int d\mathbf{p} \left\{ 1 - \frac{\hbar^2}{8} A_1^{eff} \right\} \times e^{\beta(E-H)} \quad (10)$$

with

$$A_1^{eff}(\mathbf{q}, \beta) = \frac{2}{3}\beta^2 \nabla^2 V - \frac{\beta^3}{3}(\nabla V)^2. \quad (11)$$

The momentum and time integrals then give:

$$\Delta_{avg}(\mathbf{q}, E) = \frac{1}{2\pi\hbar^2} \left\{ \int_X^\infty dz \ \text{Ai}(z) \right.$$
$$\left. + \frac{\hbar^2}{12} \nabla^2 V \left( \frac{2}{(\hbar|\nabla V|)^{2/3}} \right)^2 \text{Ai}'(X) \right\} \quad (12)$$

where Ai is the Airy function, $\text{Ai}'(z) = d\text{Ai}(z)/dz$, and X is defined as

$$X = \frac{2(V-E)}{(\hbar|\nabla V|)^{2/3}}. \quad (13)$$

Berry [11] has obtained the first term in Eq. (12) by focusing his attention on the coordinate space region close to the energy contour $V(\mathbf{q}) = E$. The second term in Eq. (12), however, is of the same order of $\hbar$ and must be included in an asymptotic expansion in $\hbar$:

(1) In the energetically allowed region we get ($X \ll 0$):

$$\Delta_{avg}(\mathbf{q}, E) = \frac{1}{2\pi\hbar^2} \left\{ 1 - \pi^{-1/2}|X|^{-3/4} \right.$$
$$\times \cos\left(\frac{2}{3}|X|^{3/2} + \frac{\pi}{4}\right)$$
$$\left. \times \left[1 - \frac{2}{3}\frac{\nabla^2 V}{(\nabla V)^2}(V-E)\right] \right\}. \quad (14)$$

The first term is the Thomas–Fermi spatial probability density.

(2) Outside of the energetically allowed region we get ($X \gg 0$):

$$\Delta_{avg}(\mathbf{q}, E) = \frac{1}{2\pi\hbar^2} \pi^{-1/2} X^{-3/4} \exp\left(-\frac{2}{3}X^{3/2} + \frac{\pi}{4}\right)$$
$$\times \left[1 - \frac{2}{3}\frac{\nabla^2 V}{(\nabla V)^2}(V-E)\right]. \quad (15)$$

### B. Periodic Orbit contribution

The oscillatory part of the spectral probability density $\Delta_{osc}(\mathbf{q}, E)$ is given by

$$\Delta_{osc}(\mathbf{q}, E) = \frac{1}{\pi\hbar} \text{Re} \int_\tau^\infty dt \ \langle \mathbf{q}|e^{-i\hat{H}t/\hbar}|\mathbf{q}\rangle e^{iEt/\hbar}. \quad (16)$$

A semiclassical expression is obtained by inserting the Van Vleck–Gutzwiller propagator in Eq. (16) and evaluating the ensuing time integral by stationary phase. We get [1,2]:

$$\Delta_{osc}(\mathbf{q}, E) = \frac{1}{\pi\hbar}\text{Re}\sum_{\text{c.t.}} \frac{1}{\sqrt{2\pi i\hbar}} \sqrt{|D|} \exp\left(\frac{i}{\hbar}S - i\frac{\pi}{2}\mu\right). \quad (17)$$

The sum is over *all* classical trajectories of energy $E$ that start at $\mathbf{q}$ and end at $\mathbf{q}$ in a nonzero time. The action $S(\mathbf{q}, \mathbf{q}, E)$ is given by

$$S(\mathbf{q}'', \mathbf{q}', E) = \int_{\mathbf{q}'}^{\mathbf{q}''} \mathbf{p}(\mathbf{q}) \, d\mathbf{q} \quad (18)$$

with $\mathbf{q}'' = \mathbf{q}' = \mathbf{q}$. The amplitude term involves the determinant

$$D(\mathbf{q}, E) = (-1)\text{Det}\begin{pmatrix} \frac{\partial^2 S}{\partial \mathbf{q}'' \partial \mathbf{q}'} & \frac{\partial^2 S}{\partial \mathbf{q}'' \partial E} \\ \frac{\partial^2 S}{\partial E \partial \mathbf{q}'} & \frac{\partial^2 S}{\partial^2 E} \end{pmatrix}\bigg|_{\mathbf{q}''=\mathbf{q}'=\mathbf{q}} \quad (19)$$

The index $\mu(\mathbf{q})$ is equal to the number of times the determinant $D$ becomes singular along the classical trajectory. To make this sum more tractable a smoothing over the spatial coordinates $\mathbf{q}$ and over the energy $E$ is done:

*a. Coordinate smoothing.* We smooth Eq. (17) over the spatial coordinates $\mathbf{q}$ and evaluate this smoothing by stationary phase. The stationary phase condition demands that the initial and final momentum of the classical trajectory be equal. Periodic orbits and orbits in their immediate vicinity therefore provide the dominant contributions to the coordinate smoothing integral. In the following we shall assume that this coordinate smoothing has been done and was sufficient to wash out the contribution of orbits not in the vicinity of periodic orbits. For a given periodic orbit it is then convenient to use a special coordinate system which moves along the trajectory [1]: the $q_1$ coordinate is chosen along the trajectory and the $q_2$ coordinate is chosen perpendicular to it. The oscillating part of the spectral probability density then becomes [2]:

$$\Delta_{osc}(\mathbf{q}, E) = \frac{2}{(2\pi\hbar)^{3/2}} \sum_{\text{p.o.}} \frac{1}{|\dot{q}_1|} \frac{1}{\sqrt{|m_{12}|}}$$
$$\times \sigma \cos\left(\frac{1}{\hbar}\left(S + \frac{1}{2}\frac{tr(\mathcal{M}) - 2}{m_{12}} q_2^2\right) - \frac{\pi}{2}\mu - \frac{\pi}{4}\right) \quad (20)$$

where $\mathcal{M}$ is the 2 x 2 submatrix of the monodromy matrix involving coordinates $q_2$ and $p_2$, and $m_{12}$ is one of its off-diagonal elements. The index $\mu(\mathbf{q})$ is given by [12]

$$\mu(\mathbf{q}) = \begin{cases} \mu_m & \text{if } \text{sign}(m_{12}) = \text{sign}(tr(\mathcal{M})) \\ \mu_m - 1 & \text{if } \text{sign}(m_{12}) = -\text{sign}(tr(\mathcal{M})) \end{cases}$$

where $\mu_m$ is the Maslov index of the orbit [13] and $\text{sign}(a) = a/|a|$. The multiplicity of the periodic orbit is denoted by $\sigma$ and is dependent on the type of orbit:

(1) $\sigma = 2$ for a libration (self-retracing orbit),

(2) $\sigma = 1$ for a rotation, where the time reversed image of a rotation is considered as being a distinct periodic orbit.

*b. Energy Smoothing.* For a chaotic system, the number of periodic orbits with period less than $\tau$ increases exponentially [14]:

$$dN \to \frac{1}{\tau}e^{\sigma\tau}d\tau \qquad \text{as} \quad \tau \to \infty$$

where $dN$ is the number of periodic orbits with periods between $\tau$ and $\tau + d\tau$ and topological entropy $\sigma$ is a measure of how chaotic the system is. In order to counter this exponential proliferation, we introduce an energy smoothing in Eq. (4):

$$\Delta(\mathbf{q}, E; \epsilon) = \sum_k f_\epsilon(E - E_k) |\langle \mathbf{q}|\psi_k\rangle|^2$$
$$= \int f_\epsilon(E - E')\Delta(\mathbf{q}, E')dE'$$
$$= \frac{1}{\pi\hbar}\text{Re}\int_0^\infty dt \tilde{f}_\epsilon(t)\langle \mathbf{q}|e^{-i\hat{H}t/\hbar}|\mathbf{q}\rangle e^{iEt/\hbar} \quad (21)$$

where $f_\epsilon(E)$ is a smoothing function and $\tilde{f}_\epsilon(t)$ is its Fourier transform. We assume that $f_\epsilon(E)$ approaches $\delta(E)$ as $\epsilon \to 0$. The energy smoothing parameter $\epsilon$ must be chosen large enough so that the periodic orbit theory converges for the number of periodic orbits available yet small enough so that we can discern details about the energy spectrum [2,7,15]. Equation (20) then becomes

$$\Delta_{osc}(\mathbf{q}, E; \epsilon) = \frac{2}{(2\pi\hbar)^{3/2}} \sum_{\text{p.o.}} \tilde{f}_\epsilon(\tau) \frac{1}{|\dot{q}_1|} \frac{1}{\sqrt{|m_{12}|}} \sigma$$
$$\times \cos\left(\frac{1}{\hbar}\left(S + \frac{1}{2}\frac{tr(\mathcal{M}) - 2}{m_{12}} q_2^2\right) - \frac{\pi}{2}\mu - \frac{\pi}{4}\right). \quad (22)$$

The sum in Eq. (22) includes periodic orbits which are repeated traversals of more fundamental periodic orbits. It is therefore convenient to rewrite this equation as a sum over these primitive periodic orbits (ppo):

$$\Delta_{osc}(\mathbf{q}, E; \epsilon) = \frac{2}{(2\pi\hbar)^{3/2}} \sum_{ppo} \sum_{n=1}^\infty \tilde{f}_\epsilon(\tau^{(n)}) \frac{1}{|\dot{q}_1|} \frac{1}{\sqrt{|m_{12}^{(n)}|}} \sigma$$
$$\times \cos\left(\frac{1}{\hbar}\left(\bar{S}^{(n)} + \frac{1}{2}\frac{tr(\mathcal{M}^{(n)}) - 2}{m_{12}^{(n)}} q_2^2\right) - \frac{\pi}{2}\mu - \frac{\pi}{4}\right) \quad (23)$$

where the superscript $n$ denotes a quantity which corresponds to the $n$-th repetition of a primitive periodic orbit and [2]:

$$\tau^{(n)} = n\tau, \qquad \bar{S}^{(n)} = n\bar{S}, \qquad \mu^{(n)} = (n-1)\mu_m + \mu$$

$$m_{12}^{(n)} = \frac{\lambda_1^n - \lambda_2^n}{\lambda_1 - \lambda_2} m_{12}, \qquad tr(\mathcal{M}^{(n)}) = \lambda_1^n + \lambda_2^n$$

The oscillating part of the semiclassical density of states, $d_{osc}(E;\epsilon)$, is obtained by taking the trace of the oscillating part of the smoothed spectral probability density:

$$d_{osc}(E;\epsilon) = \int d\mathbf{q}\, \Delta_{osc}(\mathbf{q}, E;\epsilon).$$

With

$$\int_{-\infty}^{+\infty} e^{iax^2} dx = \sqrt{\frac{\pi}{|a|}} e^{i\frac{\pi}{4} sign(a)}$$

and

$$\int dq_1 \frac{1}{|\dot{q}_1|} = \begin{cases} \tau^{(1)} & \text{for rotations} \\ \tau^{(1)}/2 & \text{for librations} \end{cases}$$

we arrive at Gutzwiller's trace formula [1]:

$$d_{osc}(E;\epsilon) = \frac{1}{\pi\hbar} \sum_{ppo} \sum_{n=1}^{\infty} \tilde{f}_\epsilon(n\tau) \frac{\tau}{\sqrt{|tr(\mathcal{M}^{(n)}) - 2|}} \\ \times \cos[n(\frac{\bar{S}}{\hbar} - \frac{\pi}{2}\mu_m)] \qquad (24)$$

where $\tau$ is the period of the underlying primitive orbit. The multiplicity of the orbits in the above formula is 1 for librations and 1 for rotations (we still consider a rotation and its time reversed image as distinct periodic orbits).

## III. PERIODIC ORBITS

The anharmonic Hamiltonian system given in Eq. (1) has been studied extensively by Baranger and Davies [5]. They found several periodic orbits using an accelerated continuation method called the Monodromy method. At energy $E = 0.001$ the main periodic orbits are the oscillations in the $x$ and $y$ directions. As the energy is increased these periodic orbits bifurcate. Hence, by following these periodic orbits and their bifurcations as the energy is increased they were able to find many periodic orbit families [5]. However there was no guarantee that all periodic orbits were found: islands of periodic orbits in the $E - \tau$ plane can appear. Shooting techniques are also impractical when the Hamiltonian system under consideration is moderately chaotic: an orbit which at some initial time is very close to a periodic orbit will be violently separated from that orbit by the latter's unstable manifold.

The method presented below guarantees that all periodic orbits with period less than a chosen maximum are found [16]. It complements other methods which are highly successful in finding periodic orbits when the Hamiltonian system is regular or slightly irregular, but break down when the system becomes moderately chaotic. We first introduce this method and then apply it to the Nelson Hamiltonian.

### A. The Surface of Section Method

Consider a two–dimensional autonomous Hamiltonian system with Hamiltonian $H(q_1, q_2, p_1, p_2)$. Classical trajectories satisfy Hamilton's equations of motion:

$$\dot{q}_i = \frac{\partial H}{\partial p_i}, \qquad \dot{p}_i = -\frac{\partial H}{\partial q_i}, \qquad i = 1, 2$$

and although the associated phase space is four–dimensional, these trajectories are restricted to the three–dimensional energy shell. A suitably chosen two–dimensional surface transversal to the phase space flow of trajectories can therefore provide a great deal of information. Such a surface is called a Poincaré surface of section [17]. The successive piercing of this surface by a trajectory produces a sequence of points $\{\ldots, x_{-1}, x_0, x_1, x_2, \ldots\}$ which defines the Poincaré map $P: x_i \mapsto x_{i+1}$. The problem of finding periodic orbits is then reduced to one of finding fixed points of the Poincaré map $P$ and of its repetitions $P^k$.

Consider a simple closed curve $\mathcal{C}$ on the surface of section and denote by $\mathcal{C}'$ the curve resulting from the Poincaré map $P$ of $\mathcal{C}$. This new curve $\mathcal{C}'$ will be greatly distorted by the stable and unstable manifolds of the periodic orbits enclosed by $\mathcal{C}$: $\mathcal{C}'$ will be stretched along the unstable manifolds and correspondingly compressed in the direction of the stable manifolds. A necessary condition for the existence of a fixed point of $P$ to be enclosed by $\mathcal{C}$ is that there must be an overlap between the area enclosed by $\mathcal{C}$ *and* the area enclosed by $\mathcal{C}'$. If there is no such overlap there cannot be a fixed point of $P$ in the area enclosed by $\mathcal{C}$ *or* by $\mathcal{C}'$. Similarly we obtain the curve $\mathcal{C}''$ as the map $P^{-1}$ of $\mathcal{C}$. Then a necessary, but not sufficient, condition for the existence of a fixed point in $\mathcal{C}$ is that there must be a common overlap between the areas enclosed by $\mathcal{C}$, $\mathcal{C}'$, and $\mathcal{C}''$.

For a chaotic system this common area will be much smaller then the original search area enclosed by $\mathcal{C}$. The Poincaré–Cartan integral invariant [17] imposes the following restriction on the projected areas enclosed by the curves $\mathcal{C}$, $\mathcal{C}'$, and $\mathcal{C}''$:

$$\oint_{\mathcal{C}} \sum_{i=1}^{2} p_i dq_i = \oint_{\mathcal{C}'} \sum_{i=1}^{2} p_i dq_i = \oint_{\mathcal{C}''} \sum_{i=1}^{2} p_i dq_i.$$

For a periodic orbit with Lyapunov exponent $\lambda$ and period $\tau$ the reduction in the original search area is of order $\exp(-2\lambda\tau)$. This method therefore works best for a substantially chaotic system, which is exactly when conventional periodic orbit finders fail. A shooting algorithm applied to the much reduced search area can then be used to zoom in on the fixed point, if it exists.

Hence by choosing an appropriate set of curves $\mathcal{C}_i$ whose enclosed areas cover the energetically allowed region of the surface of section, we can find all the fixed points of $P$. Extension of this method to find the fixed

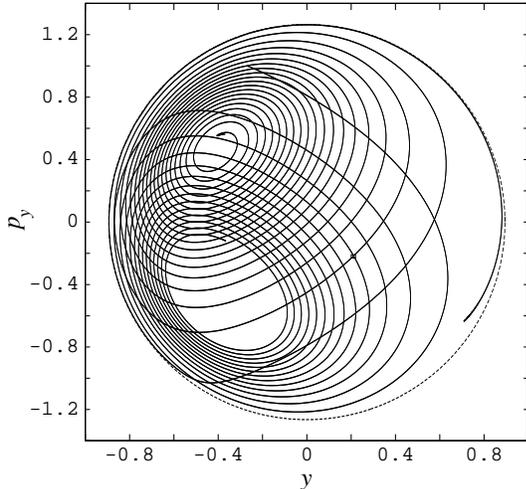

FIG. 1. The perimeter of a small square, centered at $y = 0.21$ and $p_y = -0.22$, is propagated forward and backward in time by the Poincaré map $P$ and the map $P^{-1}$, respectively. The energetically allowed region is delimited by the dashed curve, given by Eq. (25).

points of $P^k$ is immediate. We apply the above ideas to the Nelson Hamiltonian, given in Eq. (1).

### B. Periodic Orbits of the Nelson Hamiltonian

We used the Poincaré surface of section parameterized by $y$ and $p_y$, with $x = 0$ and $p_x > 0$ given by energy conservation:

$$p_x = +\sqrt{2\left(E - \frac{p_y^2}{2} - y^2\right)}$$

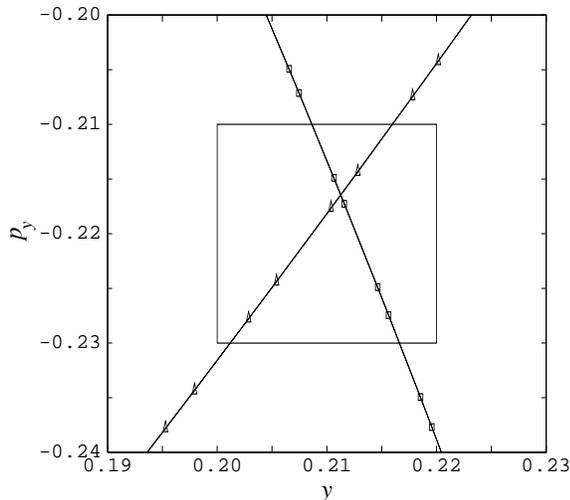

FIG. 2. An enlargement of the region of interest of Fig 1. The small boxes denote the map $P$ of some points on the square and the triangles denote the map $P^{-1}$ of some other points on the square. The reduction in search area is by a factor of $\sim 10^6$

Apart from the vertical libration along the $y$ axis, all orbits intersect this surface at distinct points. The energy was chosen to be $E = 0.8$. At these energies, the classical phase space is almost totally chaotic. We subdivided the energetically allowed region inscribed by the ellipse

$$E = \frac{p_y^2}{2} + y^2 \qquad (25)$$

into small squares of side 0.02. To efficiently propagate a side of such a square we subdivided it into unequidistant points, with a minimum separation of $10^{-7}$ and a maximum separation of $10^{-2}$. The distance between consecutive points was adjusted according to the stability of the orbits, how close the resulting propagated points were to the initial square, and how co-linear was the propagation of three consecutive points.

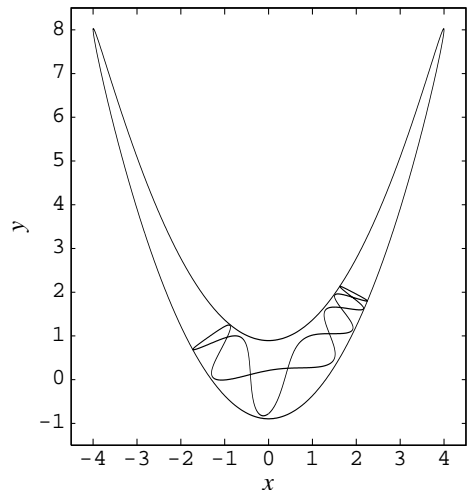

FIG. 3. The asymmetric rotation, orbit 43 in Table I, whose fixed point was found from Fig. (1) and (2).

In Fig. 1 we show such a small square and its propagation forward and backward in time by the Poincaré map $P$ and the map $P^{-1}$, respectively. In Fig. 2 we enlarged the region of interest. Note how the square as been so distorted that it effectively becomes a thin line and how much smaller the new search area is compared to the area of the initial square. In Fig. 3 we show the unstable asymmetric rotation (orbit 43 in table 1) whose fixed point was found with the help of the two previous figures. Table 1 lists the first 80 periodic orbits which were obtained by finding the fixed points of the Poincaré map $P$ and the map $P^2$. Besides its period we list what type of orbit it is, the trace of its monodromy matrix, and a point of intersection of the orbit with the surface of section. Although the maps $P$ and $P^2$ do not have any stable fixed points, we believe that there exists an $n > 2$ such that the map $P^n$ will have stable fixed points. These stable fixed points would correspond to orbits with relatively long periods [18].

TABLE I. The first 80 periodic orbits of the Nelson Hamiltonian at energy $E = 0.8$. We list whether an orbit is a symmetric libration (SL), asymmetric libration (AL), symmetric rotation (SR) or asymmetric rotation (AR). We also give its period, the trace of its Monodromy matrix and a point of intersection the orbit makes with the surface of section.

| no. | type | period | tr($\mathcal{M}$) | y | py | no. | type | period | tr($\mathcal{M}$) | y | py |
|---|---|---|---|---|---|---|---|---|---|---|---|
| 1 | SL | 4.44 | −10.6 | −0.40000 | 1.13137 | 41 | AL | 18.61 | 309.6 | 0.17951 | 0.48304 |
| 2 | SL | 6.44 | 14.7 | 0.70062 | 0.00000 | 42 | AL | 18.73 | −8735.0 | −0.44322 | 1.06915 |
| 3 | AL | 7.14 | −22.3 | 0.24646 | 1.14857 | 43 | AR | 18.74 | −1147.4 | 0.21124 | −0.21652 |
| 4 | SR | 10.51 | −136.6 | −0.88225 | 0.00000 | 44 | SR | 18.74 | −11388.1 | −0.87210 | 0.00000 |
| 5 | SL | 11.57 | 101.4 | 0.44113 | 0.00000 | 45 | SR | 18.98 | −1396.0 | −0.82779 | 0.00000 |
| 6 | AL | 11.60 | 373.0 | −0.44162 | 1.06801 | 46 | AR | 19.12 | 3540.4 | −0.29562 | 1.09236 |
| 7 | AR | 12.79 | −263.2 | −0.72413 | 0.66552 | 47 | AR | 19.20 | −2628.6 | −0.77290 | 0.53836 |
| 8 | AL | 13.66 | 525.9 | −0.41534 | 1.04915 | 48 | AL | 19.25 | 644.9 | 0.15879 | 0.18951 |
| 9 | AL | 13.66 | 175.2 | 0.35055 | 0.21226 | 49 | AR | 19.32 | −437.2 | −0.46052 | 0.90549 |
| 10 | AL | 13.69 | −359.8 | 0.65708 | 0.25149 | 50 | SR | 19.39 | −20070.8 | −0.39156 | −1.11943 |
| 11 | SL | 14.32 | 1066.8 | −0.87045 | 0.00000 | 51 | SL | 19.45 | 758.1 | 0.14911 | 0.00000 |
| 12 | AR | 14.70 | −357.9 | −0.62007 | 0.80885 | 52 | AR | 19.48 | −3191.2 | −0.85978 | 0.27340 |
| 13 | SR | 14.94 | 1584.5 | −0.32248 | −1.16141 | 53 | SR | 19.53 | 5684.7 | −0.28495 | 1.13991 |
| 14 | SR | 15.12 | −600.0 | −0.85477 | 0.00000 | 54 | AR | 19.69 | 4327.8 | −0.74676 | 0.59850 |
| 15 | AL | 15.41 | 275.6 | 0.50670 | 0.52290 | 55 | AL | 19.74 | 2428.7 | −0.71776 | 0.56791 |
| 16 | AL | 15.42 | 605.7 | −0.39170 | 1.03190 | 56 | AL | 19.78 | 546.3 | −0.32852 | 0.98436 |
| 17 | AL | 15.48 | 236.5 | 0.28249 | 0.33265 | 57 | AL | 19.83 | −6575.2 | −0.30616 | 1.07891 |
| 18 | SL | 15.79 | 344.2 | 0.26930 | 0.00000 | 58 | AR | 19.89 | 4640.7 | −0.83349 | 0.39428 |
| 19 | AL | 16.04 | −4505.3 | −0.36982 | 1.11927 | 59 | AL | 20.02 | 318.2 | 0.13613 | 0.53865 |
| 20 | AL | 16.28 | 458.2 | 0.15526 | 1.11129 | 60 | AL | 20.23 | 3518.1 | −0.81705 | 0.17590 |
| 21 | AL | 16.28 | −1123.5 | 0.21082 | −1.20070 | 61 | AR | 20.28 | −1338.1 | −0.73856 | 0.47795 |
| 22 | AR | 16.38 | −416.8 | −0.55354 | 0.86450 | 62 | AR | 20.65 | −404.6 | −0.42170 | 0.91194 |
| 23 | AL | 16.38 | 1673.9 | −0.81428 | 0.36278 | 63 | AR | 20.68 | −1815.2 | −0.80805 | 0.14789 |
| 24 | SR | 16.93 | −1619.0 | −0.88025 | 0.00000 | 64 | AL | 20.75 | 765.5 | 0.11765 | 0.24672 |
| 25 | AL | 16.99 | 626.8 | −0.36992 | 1.01575 | 65 | AL | 21.03 | 464.3 | −0.30751 | 0.96806 |
| 26 | AR | 17.05 | −900.6 | −0.81852 | 0.26813 | 66 | AL | 21.09 | 990.0 | 0.10225 | 0.07501 |
| 27 | AL | 17.11 | 281.9 | 0.22726 | 0.41700 | 67 | AL | 21.19 | 2613.6 | −0.68410 | 0.60906 |
| 28 | AR | 17.22 | 2812.4 | −0.31993 | 1.11887 | 68 | AL | 21.35 | 306.5 | 0.09494 | 0.58821 |
| 29 | AL | 17.54 | 154.3 | 0.34423 | 0.77392 | 69 | AR | 21.70 | −1475.9 | −0.70618 | 0.52906 |
| 30 | AL | 17.61 | 503.3 | 0.20776 | 0.11336 | 70 | AL | 21.82 | 4093.1 | −0.78893 | 0.27707 |
| 31 | AR | 17.65 | 3552.7 | −0.85574 | 0.27123 | 71 | AR | 21.91 | −346.6 | −0.38399 | 0.91320 |
| 32 | AR | 17.91 | −442.3 | −0.50311 | 0.89155 | 72 | SL | 22.00 | 4599.7 | −0.81335 | 0.00000 |
| 33 | AL | 17.93 | 108.7 | 0.17521 | 1.02050 | 73 | AL | 22.15 | 863.0 | 0.08168 | 0.29270 |
| 34 | SR | 17.95 | 970.8 | 0.41123 | 0.00000 | 74 | AL | 22.22 | 364.6 | −0.28530 | 0.95055 |
| 35 | AL | 18.08 | −6074.1 | −0.33694 | 1.09863 | 75 | AR | 22.22 | −2152.4 | −0.78385 | 0.24099 |
| 36 | AL | 18.14 | 5059.3 | −0.44865 | 1.07300 | 76 | SR | 22.39 | −2381.0 | −0.80242 | 0.00000 |
| 37 | AL | 18.15 | 2123.6 | −0.75958 | 0.49887 | 77 | AL | 22.55 | 2697.8 | −0.65554 | 0.63591 |
| 38 | AL | 18.37 | −1752.4 | 0.20504 | −1.20899 | 78 | AL | 22.60 | 1192.8 | 0.06291 | 0.13077 |
| 39 | AL | 18.44 | 603.2 | −0.34910 | 1.00008 | 79 | AL | 22.62 | 272.6 | 0.05395 | 0.63485 |
| 40 | SL | 18.45 | 2716.4 | −0.84063 | 0.00000 | 80 | SL | 22.74 | 1304.0 | 0.05667 | 0.00000 |

## SEMICLASSICAL RESULTS

The calculation of the exact quantal wave functions and energies were obtained by matrix diagonalization. We used the basis $\phi_m^{(x)}(x)\phi_n^{(y)}(y - x^2/2)$, where $\phi_m^{(x)}$ and $\phi_n^{(y)}$ are harmonic oscillator wave functions appropriate for low energies. The matrix elements of the Nelson Hamiltonian in this basis are given in Appendix A. We set $\hbar = 0.05$ and used 240 oscillator states for the x direction and 26 for the y direction. The maximal number of basis states in the x direction and the y direction were chosen in conjunction with the value of $\hbar$ in order to get the best possible convergence of the eigenvalues up $E \simeq 0.8$. At these energies the classical phase space is for all practical purposes totally chaotic. The accuracy of this basis was further tested against the average semiclassical spectral staircase $N_{avg}(E)$, which is given by (see Appendix B)

$$N_{avg}(E) = \frac{1}{2\sqrt{2\mu}} \frac{E^2}{\hbar^2} \left( 1 - \frac{\hbar^2}{12} \left[ \frac{\mu + 2}{E^2} + \frac{2}{\mu E} \right] + O(\hbar^4) \right). \quad (26)$$

The first term is the usual Thomas–Fermi expression for the staircase. The inclusion of corrections to the Thomas–

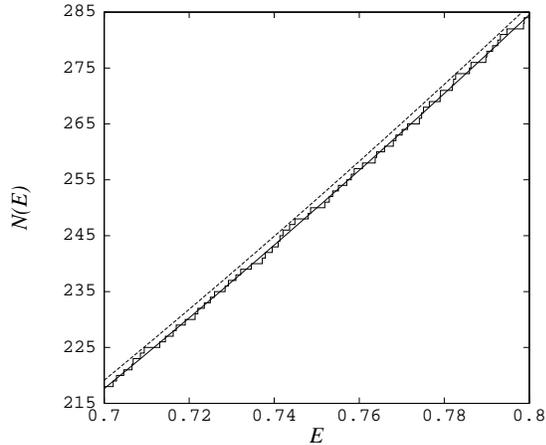

FIG. 4. Comparison of the average semiclassical staircase, and the one obtained by diagonalization. The dashed line is the Thomas-Fermi staircase.

Fermi staircase is essential to achieve a proper fit with the exact spectral staircase. This is shown in Fig. 4 for the energy interval $0.7 \leq E \leq 0.8$. The agreement between the semiclassical prediction for the average staircase $N_{avg}(E)$ and the exact staircase is excellent. In Fig. 5 we show the oscillating part of the staircase function $N_{osc}(E) = N(E) - N_{avg}(E)$. The sudden dip in $N_{osc}(E)$ around $E \simeq 0.83$ indicates the failure of the diagonalization for higher energies.

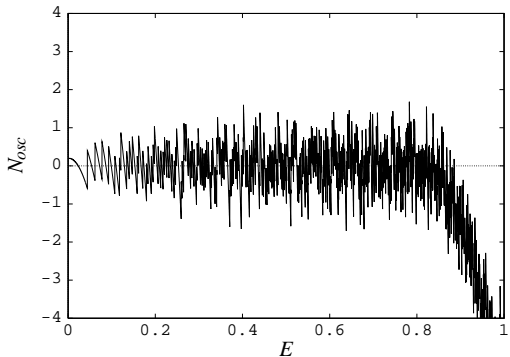

FIG. 5. The oscillatory part of the spectral staircase, $N_{osc}(E) = N(E) - N_{avg}(E)$. The sudden dip indicates the failure of the diagonalization for $E \simeq 0.83$.

### C. Density of States

Gutzwiller's trace formula, Eq. (24), provides the means of calculating semiclassically the oscillating part of the density of states $d_{osc}(E; \epsilon)$. We chose the following form for the energy smoothing:

$$f_\epsilon(E) = \frac{1}{\sqrt{2\pi\epsilon^2}} e^{-E^2/2\epsilon^2}.$$

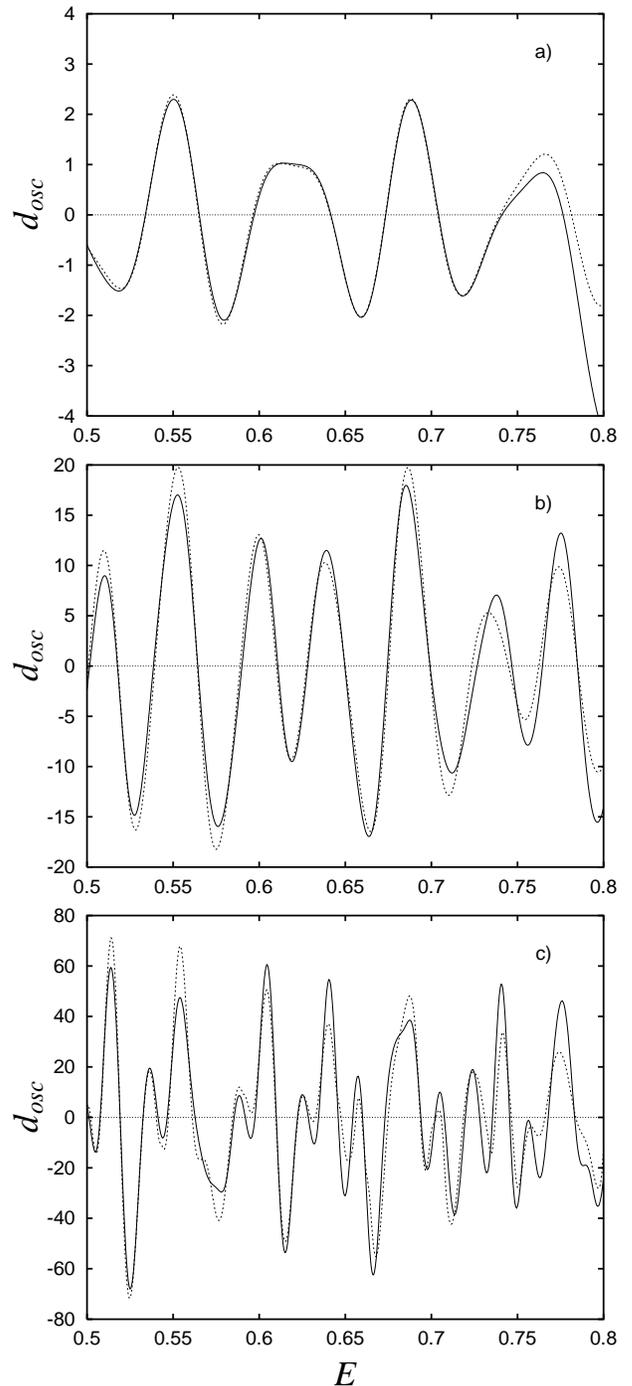

FIG. 6. The oscillating part of the smoothed energy level density, $d_{osc}(E; \epsilon)$, for three different energy smoothing parameters $\epsilon$. The dashed lines represent the semiclassical results while the solid lines display the exact results. The average energy level density at $E = 0.65$ is $\simeq 580$. Top: $\epsilon = 0.02$, and the first 2 periodic orbits in Table I contribute noticeably to the sum in Gutzwiller's trace formula. Middle: $\epsilon = 0.01$, and the first 5 periodic orbits contribute. Bottom: $\epsilon = 0.005$, and the first 60 periodic orbits contribute.

The damping of the long orbits in Gutzwiller's trace for-

mula is then given by

$$\tilde{f}_\epsilon(\tau) = e^{-\epsilon^2\tau^2/2\hbar^2}.$$

We applied Gutzwiller's trace formula with different values of the energy smoothing parameter $\epsilon$. In Fig. 6 we show our semiclassical results (dashed lines) and compare them with the exact calculations (solid lines). We took $\epsilon = 0.02$ in Fig. 6a. Only the first 2 periodic orbits listed in Table 1 contribute noticeably to $d_{osc}(E;\epsilon)$. Since the average semiclassical density of states at $E = 0.65$ is $\simeq 580$, the maximum oscillations in the density of states is about 0.4% of the average value. The agreement between the exact and semiclassical results is excellent up to energy $E = 0.75$. It is the failure of the numerical diagonalization for energies above $E = 0.83$ that is causing the exact results for $d_{osc}(E;\epsilon)$ to diverge after $E = 0.75$. Note that the number of states in an energy range $[E - \epsilon, E + \epsilon]$ is about 24 for $\epsilon = 0.02$. Thus Gutzwiller's trace formula only describes the gross features of the clustering of the eigenvalues for this value of $\epsilon$, but does so extremely well.

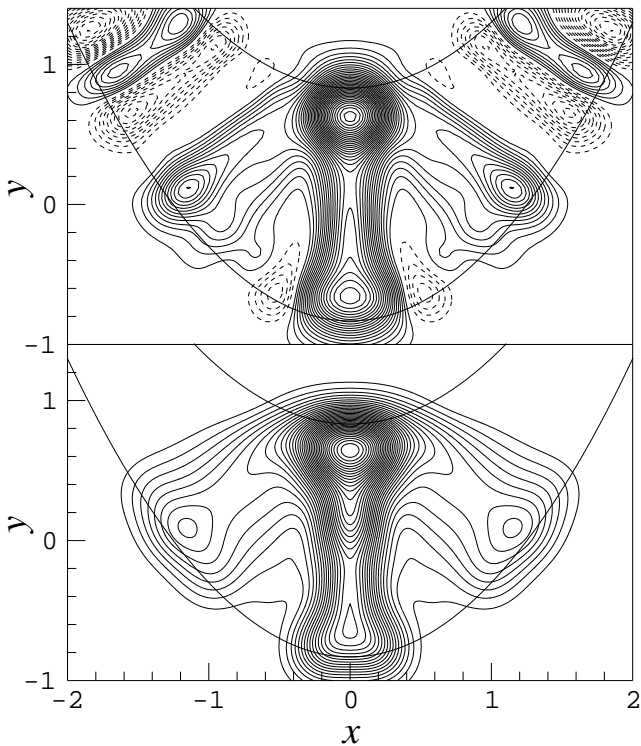

FIG. 7. The oscillating part, $\Delta_{osc}$, of the energy and space smoothed coordinate space density. Energy smoothing parameter $\epsilon = 0.01$. Top: exact quantal. Bottom: Eq. (23) with 5 orbits. Solid contours are positive, dashed are negative. The contour spacing is 0.5. The Thomas–Fermi density is 63.7. The energy is $E = 0.692$.

In Fig. 6b we took $\epsilon = 0.01$. In this case the first 5 periodic orbits contribute significantly and the maximum oscillations are about 3% of the average value and there are about 12 states within an energy range $[E - \epsilon, E + \epsilon]$. The agreement is still very good. However, there are some small discrepancies and the inclusion of more periodic orbits do not help. This is also the case in Fig. 6c where $\epsilon = 0.005$. Only the first 60 periodic orbits contribute and now the maximum oscillations are 13% of the average value and about 6 states are in an energy range $[E - \epsilon, E + \epsilon]$. Since Gutzwiller's trace formula is the first term in an expansion in powers of $\hbar$ of $d_{osc}(E;\epsilon)$, we believe that higher order corrections are needed to get better agreement between the exact and semiclassical results when $\epsilon$ becomes smaller.

### D. Spectral Probability Density

Bogomolny's formula, Eq. (23), provides us with the means of calculating semiclassically the oscillatory part of the spectral probability density $\Delta_{osc}(\mathbf{q}, E;\epsilon)$. In Figs. 7, 8, and 9 we show our semiclassical results (bottom half of figures) and compare them with the exact calculations (top half of figures). The solid contours indicate when $\Delta_{osc}(\mathbf{q}, E;\epsilon)$ is positive and the dashed contours when $\Delta_{osc}(\mathbf{q}, E;\epsilon)$ is negative. The contour spacings is 0.5 in all Figs. 7, 8, and 9.

The energy smoothing parameter was set to $\epsilon = 0.01$ and we found that only the first 5 periodic orbits contribute significantly. We also performed a gaussian smoothing over the coordinate space with

$$\frac{1}{\pi b^2} e^{-(q_x^2+q_y^2)/b^2}.$$

We took $b = 0.2$, which is sufficient to wash out the contribution of most non–periodic orbits in the central region of the potential. However this value of the smoothing parameter $b$ is too small for the "arms" region of the potential. The nodal structure seen up the arms of the potential can easily be described using WKB expressions for the wavefunctions since the classical motion up the arms is adiabatically separable. We therefore concentrated our attention to the central part of the potential. The chosen value of $b$ is a compromise between choosing $b$ small enough to see details in the central region and large enough to eliminate the contribution of non–periodic orbits to $\Delta_{osc}(\mathbf{q}, E;\epsilon)$. It is interesting to note that the gaussian coordinate smoothing of the spectral probability density is equivalent to the coordinate space projection of the coherent state representation of the spectral operator [19].

The average part of the coordinate smoothed and energy smoothed spectral probability density was calculated using both Eq. (12) and Eq. (C1) from Appendix C. It was found that Eq. (C1) sufficed for the present accuracy of the calculations and was much faster. The Thomas–Fermi density is 63.7. The maximum oscillations in $\Delta_{osc}(\mathbf{q}, E;\epsilon)$ were around 10 or about 15% of $\Delta_{avg}(\mathbf{q}, E;\epsilon)$.

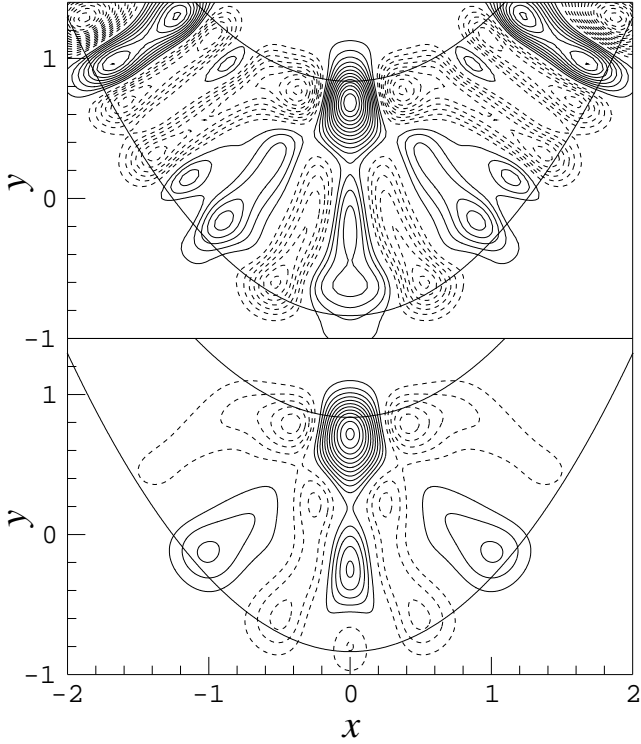

FIG. 8. Same as Fig. 7 for $E = 0.700$.

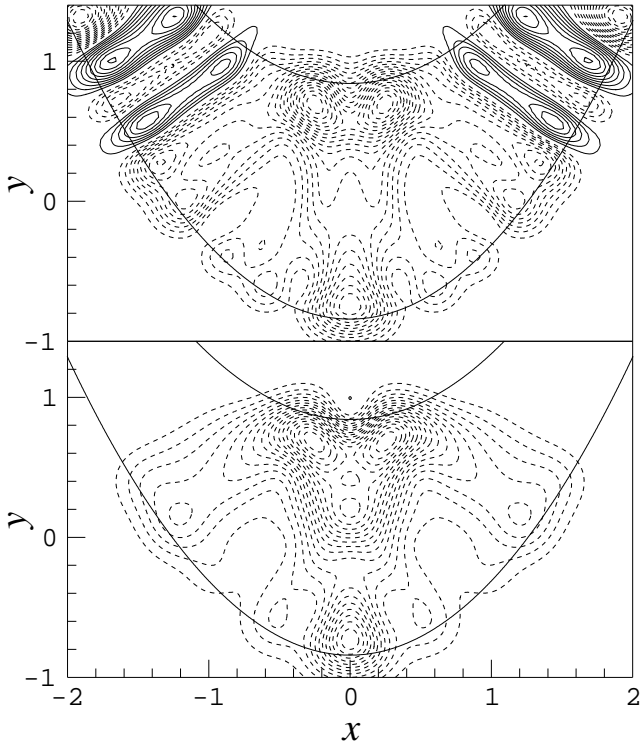

FIG. 9. Same as Fig. 7 for $E = 0.708$.

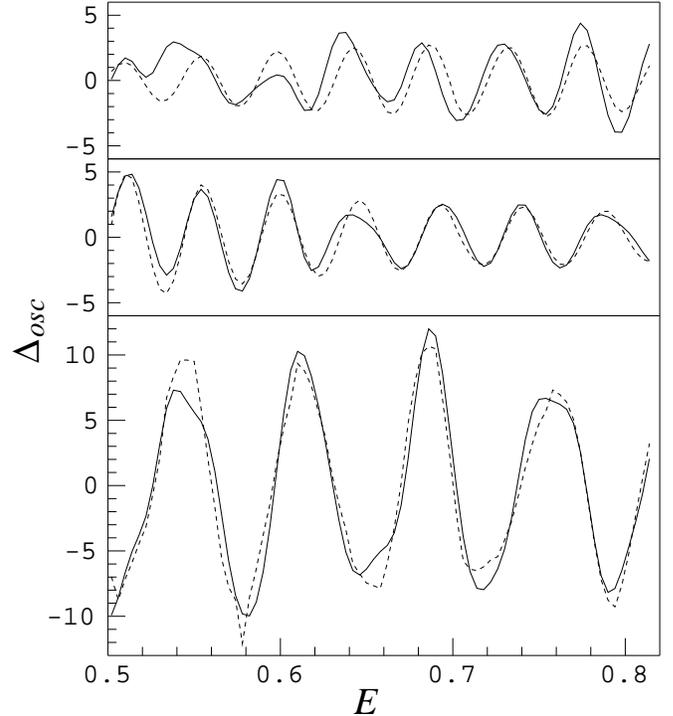

FIG. 10. The exact scar strength $\Delta$ (full line) and the semiclassical one calculated with the first 5 orbits in Table I (dashed), as a function of energy, at three points of coordinate space. Bottom: $x = 0$, $y = -0.65$, emphasizing orbit no. 1. Middle: $x = 0.8$, $y = 0$, emphasizing orbit no. 2. Top: $x = 0.75$, $y = 0.75$, emphasizing orbit no. 3.

The qualitative agreement between the exact quantal and the semiclassical pictures is very good. The main trends are well represented in all of our comparisons. The average density of states at $E = 0.70$ is $\simeq 626$. Hence there is a shift by about 5 eigenstates between the calculations presented in Fig. 7 and Fig. 8 and similarly between Fig. 8 and Fig. 9. Since the chosen energy smoothing parameter ($\epsilon = 0.01$) is relatively large, we were surprised that the drastic changes in $\Delta_{osc}(\mathbf{q}, E; \epsilon)$ are as well represented by the semiclassical calculations. In Fig. 10 we show $\Delta_{osc}(\mathbf{q}, E; \epsilon)$ for three points in coordinate space for $0.5 \leq E \leq 0.8$. The three points are chosen so as to emphasize separately the contribution of the three lowest periodic orbits. Again the agreement is very reasonable considering the approximations.

## IV. CONCLUSION

We showed that periodic orbits play a central role in the semiclassical description of a physical system. The method we used to find the periodic orbits guaranteed that we had all the periodic orbits up to a chosen maximum period. This allowed us to apply Gutzwiller's trace formula and Bogomolny's formula to a generic Hamiltonian system.

In our calculations of the oscillating part of the density of states, we found that when the energy smoothing parameter was large enough Gutzwiller's trace formula worked extremely well. Not only did we get a very accurate semiclassical description of the oscillations in the density of states, we did so with little numerical effort since only a few periodic orbits were needed. Also, exact quantum calculations become numerically infeasible for higher energies. Gutzwiller's trace formula then is invaluable in describing the coarse oscillations in the density of states. However when we reduced the energy smoothing parameter so as to obtain more detailed information about the spectrum, we found that there was a discrepancy between the amplitudes of the semiclassical and exact quantal calculations. These discrepancies could not be reduced by including more periodic orbits in Gutzwiller's trace formula. We noted that because the amplitude of the oscillations relative to the average density of states was not small, asymptotic corrections in $\hbar$ to Gutzwiller's trace formula are needed.

Similar calculations of the oscillatory part of the spectral probability density using Bogomolny's formula were done. We found that there was very good qualitative agreement between the exact quantal and the semiclassical calculations over a wide energy range. Since the maximum oscillations in $\Delta_{osc}(\mathbf{q}, E; \epsilon)$ were around 10 or about 15% of $\Delta_{avg}(\mathbf{q}, E; \epsilon)$, better quantitative agreement will only be possible if asymptotic corrections in $\hbar$ are included.

## V. ACKNOWLEDGEMENTS

I would like to thank Michel Baranger for many helpful discussions. This work was supported in part by funds provided by the U. S. Department of Energy (D.O.E.) under contract #DE-AC02-76ER03069.

## APPENDIX A: DIAGONALIZATION

The eigenvalues and the corresponding eigenstates of the Nelson Hamiltonian $\hat{H}$ were obtained by diagonalization in a displaced harmonic oscillator basis with the spatial coordinates $\xi_1$ and $\xi_2$:

$$\xi_1 = x, \qquad \xi_2 = y - \frac{x^2}{2}.$$

The coordinate transformation from $\{x, y\}$ to $\{\xi_1, \xi_2\}$ is non-orthogonal. A simple way to make this coordinate transformation is to write Schrödinger's equation in Cartesian coordinates,

$$\left[ -\frac{\hbar^2}{2} \left\{ \frac{\partial^2}{\partial x^2} + \frac{\partial^2}{\partial y^2} \right\} + \frac{\mu}{2} x^2 + \left( y - \frac{x^2}{2} \right)^2 \right] \psi(x, y)$$
$$= E\psi(x, y)$$

and then to make the change of variables from $\{x, y\}$ to $\{\xi_1, \xi_2\}$:

$$\left[ -\frac{\hbar^2}{2} \left\{ \frac{\partial^2}{\partial \xi_1^2} + (1 + \xi_1^2) \frac{\partial^2}{\partial \xi_2^2} - (2\xi_1 \frac{\partial}{\partial \xi_1} + 1) \frac{\partial}{\partial \xi_2} \right\} nonumber \right. \quad (A1)$$

$$\left. + \frac{\mu}{2} \xi_1^2 + \xi_2^2 \right] \psi(\xi_1, \xi_2) = E\psi(\xi_1, \xi_2) \quad (A2)$$

The matrix elements of the Hamiltonian in the harmonic oscillator basis with the variables $\xi_1$ and $\xi_2$ are obtained by making the identification

$$p_1 = \frac{\hbar}{i} \frac{\partial}{\partial \xi_1} \qquad p_2 = \frac{\hbar}{i} \frac{\partial}{\partial \xi_2}$$

and rewriting Eq. (A2) as

$$\left[ \frac{\hat{p}_1^2}{2} + (1 + \hat{\xi}_1^2) \frac{\hat{p}_2^2}{2} - (\hat{\xi}_1 \hat{p}_1 - \frac{i\hbar}{2}) \hat{p}_2 + \frac{\mu}{2} \hat{\xi}_1^2 + \hat{\xi}_2^2 \right] \psi$$
$$= E\psi$$

With $w_1 = \sqrt{0.1}$ and $w_2 = \sqrt{2}$,

$$\hat{p}_1 = \sqrt{\frac{\hbar w_1}{2}} i(\hat{a}_1^\dagger - \hat{a}_1) \qquad \hat{\xi}_1 = \sqrt{\frac{\hbar}{2w_1}}(\hat{a}_1^\dagger + \hat{a}_1)$$

$$\hat{p}_2 = \sqrt{\frac{\hbar w_2}{2}} i(\hat{a}_2^\dagger - \hat{a}_2) \qquad \hat{\xi}_2 = \sqrt{\frac{\hbar}{2w_2}}(\hat{a}_2^\dagger + \hat{a}_2)$$

the desired matrix elements of $\hat{H}$ in the basis $|n_1 n_2\rangle = |n_1\rangle_{\xi_1} |n_2\rangle_{\xi_2}$ are obtained:

$$\langle n_1 n_2 | \hat{H} | n_1 n_2 \rangle = \hbar\omega_1(n_1 + \frac{1}{2}) + \hbar\omega_2(n_2 + \frac{1}{2})$$
$$+ (\frac{\hbar}{2\omega_1})\hbar\omega_2(n_1 + \frac{1}{2})(n_2 + \frac{1}{2})$$

$$\langle n_1 n_2 | \hat{H} | n_1 \ n_2+2 \rangle = -\frac{\hbar^2 \omega_2}{4\omega_1}(n_1 + \frac{1}{2})\sqrt{(n_2+1)(n_2+2)}$$

$$\langle n_1 n_2 | \hat{H} | n_1+2 \ n_2 \rangle = \frac{\hbar^2 \omega_2}{4\omega_1}(n_2 + \frac{1}{2})\sqrt{(n_1+1)(n_1+2)}$$

$$\langle n_1 n_2 | \hat{H} | n_1+2 \ n_2+1 \rangle = \frac{\hbar}{2}\sqrt{\frac{\hbar\omega_2}{2}}\sqrt{(n_2+1)(n_1+1)(n_1+2)}$$

$$\langle n_1 n_2 | \hat{H} | n_1+2 \ n_2-1 \rangle = -\frac{\hbar}{2}\sqrt{\frac{\hbar\omega_2}{2}}\sqrt{n_2(n_1+1)(n_1+2)}$$

$$\langle n_1 n_2 | \hat{H} | n_1+2 \ n_2+2 \rangle = -\frac{\hbar^2 \omega_2}{8\omega_1}\sqrt{(n_2+1)(n_2+2)(n_1+1)(n_1+2)}$$

$$\langle n_1 n_2 | \hat{H} | n_1+2 \ n_2-2 \rangle = -\frac{\hbar^2 \omega_2}{8\omega_1}\sqrt{n_2(n_2-1)(n_1+1)(n_1+2)}$$

## APPENDIX B: AVERAGE DENSITY OF STATES

The Thomas–Fermi approximation for the staircase $N(E) = \sum_i \theta(E - E_i)$ is the number of Planck cells contained in the phase space volume enclosed by the energy shell. For the Nelson Hamiltonian,

$$N_{avg}(E) = \frac{1}{(2\pi\hbar)^2} \int d\mathbf{q} d\mathbf{p} \, \Theta\left(E - \frac{p_x^2}{2} - \frac{p_y^2}{2} - \frac{\mu}{2}x^2 - (y - \frac{x^2}{2})^2\right)$$

The set of transformations

$$x = \sqrt{\frac{2}{\mu}} z_1, \quad y - \frac{x^2}{2} = z_2, \quad p_x = \sqrt{2} z_3, \quad p_y = \sqrt{2} z_4 \tag{B1}$$

gives

$$N_{avg}(E) = 2\sqrt{\frac{2}{\mu}} \frac{1}{(2\pi\hbar)^2} \int d^4z \, \Theta\left(E - z_1^2 - z_2^2 - z_3^2 - z_4^2\right).$$

The above integral is the volume of a four dimensional sphere of radius $\sqrt{E}$. Therefore

$$N_{avg}(E) = \frac{1}{2\sqrt{2\mu}} \frac{E^2}{\hbar^2}.$$

To obtain corrections to $N_{avg}(E)$, we define $Z(\beta)$ as the trace of the evolution operator. From Eqs. (8) and (11) we get for the Nelson Hamiltonian, up to $O(\hbar^3)$:

$$Z(\beta) \simeq \frac{1}{(2\pi\hbar)^2} \int d\mathbf{q} d\mathbf{p} \, e^{-\beta\left[\frac{p_x^2}{2} + \frac{p_y^2}{2} + \frac{\mu x^2}{2} + \left(y - \frac{x^2}{2}\right)^2\right]} \times \left(1 - \frac{\hbar^2}{8} A_1^{eff}\right)$$

with

$$A_1^{eff}(x, y, \beta) = \frac{2}{3}\beta^2\left\{(\mu + 2) + 2x^2 - 2\left(y - \frac{x^2}{2}\right)\right\}$$
$$- \frac{\beta^3}{3}\left\{x^2\left[\mu^2 - 4\mu\left(y - \frac{x^2}{2}\right) + 4\mu\left(y - \frac{x^2}{2}\right)^2\right]\right.$$
$$\left. + 4\left(y - \frac{x^2}{2}\right)^2\right\}$$

The set of transformations (B1) gives

$$Z(\beta) = 2\sqrt{\frac{2}{\mu}} \frac{1}{(2\pi\hbar)^2} \int dz_1 dz_2 dz_3 dz_4 \, e^{-\beta\left[z_1^2 + z_2^2 + z_3^2 + z_4^2\right]}$$
$$\times \left\{1 - \frac{\hbar^2}{12}\left[\beta^2\left((\mu + 2) + \frac{4z_1^2}{\mu}\right)\right.\right.$$
$$\left.\left. - \frac{\beta^3}{3}\left[\frac{2z_1^2}{\mu}(\mu^2 + 4z_2^2) + 4z_2^2\right]\right]\right\}$$

which simplifies to

$$Z(\beta) = \sqrt{\frac{2}{\mu}} \frac{1}{2\hbar^2} \left(\frac{1}{\beta^2} - \frac{\hbar^2}{24}\left[(\mu + 2) + \frac{2}{\beta\mu}\right]\right)$$

Taking the inverse Fourier transform of $Z(\beta)/\beta$ we get

$$N_{avg}(E) = \frac{1}{2\sqrt{2\mu}} \frac{E^2}{\hbar^2} \left(1 - \frac{\hbar^2}{12}\left[\frac{\mu + 2}{E^2} + \frac{2}{\mu E}\right] + O(\hbar^4)\right)$$

Finally the derivative of $N_{avg}(E)$ with respect to $E$ gives the average level density $d_{avg}(E)$:

$$d_{avg}(E) = \frac{1}{\sqrt{2\mu}} \frac{E}{\hbar^2} \left(1 - \frac{\hbar^2}{12} \frac{1}{\mu E} + O(\hbar^4)\right)$$

## APPENDIX C: AVERAGE SPECTRAL PROBABILITY DENSITY

We found during our numerical work on the spatial wavefunctions that coordinate and energy smoothing of the Thomas–Fermi spatial density was sufficient to reproduce the average spectral probability density. The Thomas–Fermi spatial density is

$$\Delta_{\mathrm{TF}}(\mathbf{q}, E) = \begin{cases} \frac{1}{2\pi\hbar^2} & \text{if} \quad V(\mathbf{q}) \leq E \\ 0 & \text{otherwise} \end{cases}$$

With a gaussian smoothing in energy, we get

$$\Delta_{avg}(\mathbf{q}, E) = \frac{1}{2\pi\hbar^2} \int_{-\infty}^{\infty} f_\epsilon(E' - E) dE'$$
$$= \frac{1}{2\pi\hbar^2} \frac{1}{\sqrt{2\pi\epsilon^2}} \int_{V(\mathbf{q})}^{\infty} \exp\left[-(E' - E)^2/2\epsilon^2\right] dE'$$
$$= \frac{1}{2\pi\hbar^2} \frac{1}{2} \mathrm{erfc}\left(\frac{V - E}{\epsilon\sqrt{2}}\right) \tag{C1}$$

where erfc is the complementary error function. The coordinate smoothing was then done numerically.